\documentclass{PoS}
\usepackage{amsmath}
\usepackage{subfigure}

\title{The \boldmath$B\to D^\ast\ell\nu$ semileptonic decay at nonzero recoil and its implications for $\left|V_{cb}\right|$ and $R(D^\ast)$}

\ShortTitle{\boldmath$B\to D^\ast\ell\nu$ at nonzero recoil}

\author{\speaker{Alejandro Vaquero Avil\'es-Casco}\\
        Department of Physics and Astronomy, University of Utah, Salt Lake City, UT 84112-0830, USA\\
        E-mail: \email{alexvaq@physics.utah.edu}}

\author{Carleton DeTar\\
        Department of Physics and Astronomy, University of Utah, Salt Lake City, UT 84112-0830, USA\\
        E-mail: \email{detar@physics.utah.edu}}

\author{Aida X. El-Khadra\\
        Department of Physics, University of Illinois, Urbana, IL 61801-3080, USA and \\
        Fermi National Accelerator Laboratory, Batavia, IL 60510-5011, USA\\
        E-mail: \email{axk@illinois.edu}}

\author{Andreas S. Kronfeld\\
        Fermi National Accelerator Laboratory, Batavia, IL 60510-5011, USA\\
        E-mail: \email{ask@fnal.gov}}

\author{Jack Laiho\\
        Department of Physics, Syracuse University, Syracuse, NY 13244-1130, USA\\
        E-mail: \email{jwlaiho@syr.edu}}

\author{Ruth S. Van de Water\\
        Fermi National Accelerator Laboratory, Batavia, IL 60510-5011, USA\\
        E-mail: \email{ruthv@fnal.gov}}

\author{(Fermilab Lattice and MILC Collaborations)}

\abstract{We present nearly final results from our analysis of the form factors for $B\to D^\ast\ell\nu$ decay at nonzero recoil. Our analysis includes 15 MILC asqtad ensembles with $N_f=2+1$ flavors of sea
          quarks and lattice spacings ranging from $a\approx0.15$ fm down to $0.045$ fm. The valence light quarks employ the asqtad action, whereas the $b$ and $c$ quarks are treated using the Fermilab action.
          We discuss the impact that our results will have on $\left|V_{cb}\right|$ and $R(D^\ast)$.}

\FullConference{37th International Symposium on Lattice Field Theory - Lattice2019 \\
		16-22 June 2019 \\
		Wuhan, China}

\begin{document}

\section{Introduction}
The CKM~\cite{Cabibbo:1963yz, Kobayashi:1973fv} matrix element $\left|V_{cb}\right|$ has been lately subjected to quite a controversial discussion whose origin, the tension between determinations based on
inclusive and exclusive decay analyses, dates back several years. As a reminder, the determination of $\left|V_{cb}\right|$ from experimental measurements of exclusive decay rates requires, as an input,
theoretical calculations of the decay amplitudes, but so far these calculations have been carried out only at zero recoil. During the last two years, a few articles that pointed towards a possible resolution of
the inclusive-exclusive tension were published~\cite{BIGI2017441, GRINSTEIN2017359}. The argument is simple: at fault is the parameterization of the shape of the form factors used to extrapolate the
experimental decay rate to zero recoil in the exclusive determinations. The more restrictive Caprini-Lellouch-Neubert (CLN) parametrization~\cite{Caprini:1997mu} gave results that disagreed with the inclusive
value, whereas the more general Boyd-Grinstein-Lebed (BGL) parametrization~\cite{Boyd:1997kz} eliminated the tension. As a result, the latest PDG review~\cite{Tanabashi:2018oca} (2018) reports no tension
between inclusive and exclusive determinations of $\left|V_{cb}\right|$. The situation changed when the experimental collaborations decided to carry out an analysis of their existing data using the BGL
parametrization, in order to check whether the inclusive-exclusive tension indeed had disappeared. Both Belle~\cite{BelleUntagged} and BaBar~\cite{BaBar} collaborations reported that the discrepancy was still
there, regardless of the parametrization used, and the current consensus is that we do not understand why the exclusive determination differs from the inclusive one. Another important tension between the
Standard Model and experiment related to this process is in $R(D)$ and $R(D^\ast)$. In the $R(D)-R(D^\ast)$ plane the tension between theoretical predictions and experiment is $\approx 3\sigma$
\cite{Amhis:2019ckw}. There exist several theoretical determinations of $R(D)$ coming from lattice QCD~\cite{Lattice:2015rga,Na:2015kha}. In contrast, none of the determinations of $R(D^\ast)$ are based on
unquenched lattice QCD results~\cite{Fajfer:2012vx,Bernlochner:2017jka,Bigi:2017jbd,Jaiswal:2017rve}.

This work aims to address these issues by performing a complete analysis of the $B\rightarrow D^\ast\ell\nu$ form factors at non-zero recoil on the lattice. Here we present a preliminary result for the form
factors, whose normalization is blinded by an overall multiplicative factor. The results presented here are still preliminary, but many aspects of the analysis procedure have already been cross checked. There
are also other ongoing efforts of lattice QCD calculations of the same quantities using a different regularization and independent ensembles~\cite{Kaneko:2018mcr,JLQCD}.

\section{Notation and definitions}
This decay is mediated by the $V$-$A$ weak currents,
\begin{align}
\frac{\left\langle D^\ast(p_{D^\ast},\epsilon_\nu)|\mathcal{V}^\mu|B(0)\right\rangle}{2\sqrt{M_{D^\ast}M_B}} =& \frac{1}{2}\epsilon^*_\nu\varepsilon^\mu_{\nu\rho\sigma} v_{D^\ast}^\rho v_B^\sigma h_V(w), \\
\frac{\left\langle D^\ast(p_{D^\ast},\epsilon_\nu)|\mathcal{A}^\mu|B(0)\right\rangle}{2\sqrt{M_{D^\ast}M_B}} =& \frac{i}{2}\epsilon^*_\nu\left[g^{\mu\nu}(1+w)h_{A1}(w) - v_B^\nu\left(v_B^\mu h_{A_2}(w) + v_{D^\ast}^\mu h_{A_3}(w)\right)\right],
\end{align}
where $\mathcal{A}^\mu$ and $\mathcal{V}^\mu$ are the continuum axial and vector currents, $M_X$ is the mass of the meson $X$, $p_X$ its momentum and $v_X$ its four-velocity, $w=v_B^\mu v_{D^\ast}^\mu$ is the
recoil parameter, and $\epsilon_\nu$ is the polarization of the $D^\ast$, The different form factors $h_X(w)$ are motivated by Heavy Quark Effective Theory (HQET), and they enter in the definition of the decay
amplitude $\mathcal{F}(w)$
\begin{equation}
\chi(w)\mathcal{F}(w) = \frac{1 - 2wr + r^2}{12M_{D^\ast}M_B(1-r)^2}\left[H_+^2(w) + H_0^2(w) + H_-^2(w)\right],
\end{equation}
through the helicity amplitudes $H_\pm$, $H_0$ and $H_S$, the latter contributing only when the mass of the lepton is not negligible. $\chi(w)$ is just a kinematic factor.
The quantity measured in experiments is the differential decay rate, defined as
\begin{equation}
\frac{d\Gamma}{dw}\left(\bar{B}\rightarrow D^\ast\ell\bar{\nu}_\ell\right) = \frac{G_F^2 m_B^5}{48\pi^2}(w^2-1)^{\frac{1}{2}}P(w)\left|\eta_{ew}\right|^2 \left|\mathcal{F}(w)\right|^2 \left|V_{cb}\right|^2,
\end{equation}
The aim of this work is to calculate the $h_X(w)$ form factors as a function of the recoil parameter.

\section{Simulation details}
Our simulation employs 15 ensembles of $2+1$ asqtad sea quarks, whose strange quark masses are tuned to the physical value. The heavy quarks appear only in the valence sector and use the clover action with
the Fermilab interpretation. Their masses are tuned to reproduce the physical masses of the $B_s$ and the $D_s$ mesons on each ensemble. The lattice spacings of our ensembles range from $a=0.15$ in the
coarsest case to $a=0.045$ fm in the finest ensemble, and the pion masses start at $m_\pi\approx 550$ MeV down to $m_\pi\approx 180$ MeV for the lightest ensemble. There are no ensembles at the physical pion
mass.

\begin{figure}[h!]
  \centering
  \includegraphics[angle=0,width=6.0cm]{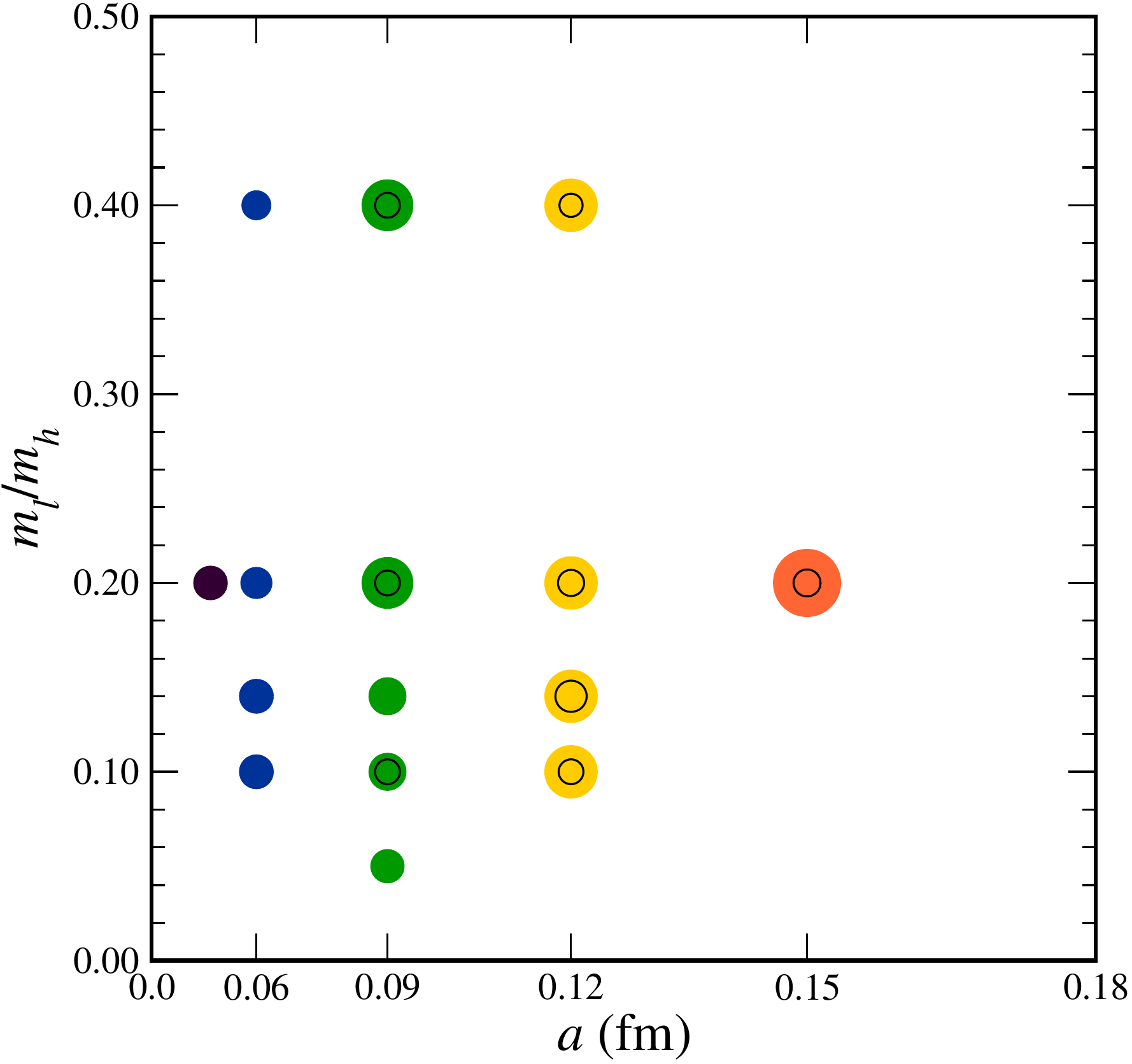}
  \caption{Lists of available ensembles. On the $x$ axis the ensebmles are ordered by their lattice spacing, and the $y$ axis show the ratio between the light and the strange quark mass. The physical ratio
           stands at $\approx 1/27$. The size of each circle gives information on the size of the ensemble.\label{eList}}
\end{figure}

The amount of data available and the characteristics of each ensemble are plotted in Fig.~\ref{eList}, where the size of the circle gives information about the statistics on each ensemble.

Calculations were done at two different momenta ${\bf p}^2 = \left(2\pi/L\right)^2, \left(4\pi/L\right)^2$, where $L$ is the size of the lattice, except for the $h_{A_1}$ form factor that also included
a direct zero momentum calculation. 

\section{Lattice results}
After calculating the relevant ratios of correlation functions required to compute the form factors, the matching factors $\rho_{V,A}$, which are computed perturbatively, are applied~\cite{Harada:2001fj}.
Then a correction is added to fix small mistunings in the value of the heavy quark masses and the form factors are extracted. The non-perturbative piece of the renormalization factors is explicitly cancelled
in the ratios we construct. We introduce a blinding procedure at the level of the matching factors. All the $\rho$ factors are multiplied by a single number, close to one. As a result, all the results shown
hereafter are blinded.
\begin{figure}[h]
  \centering
  \subfigure[$h_V(w)$ form factor.]
            {\includegraphics[width=0.40\linewidth,angle=0]{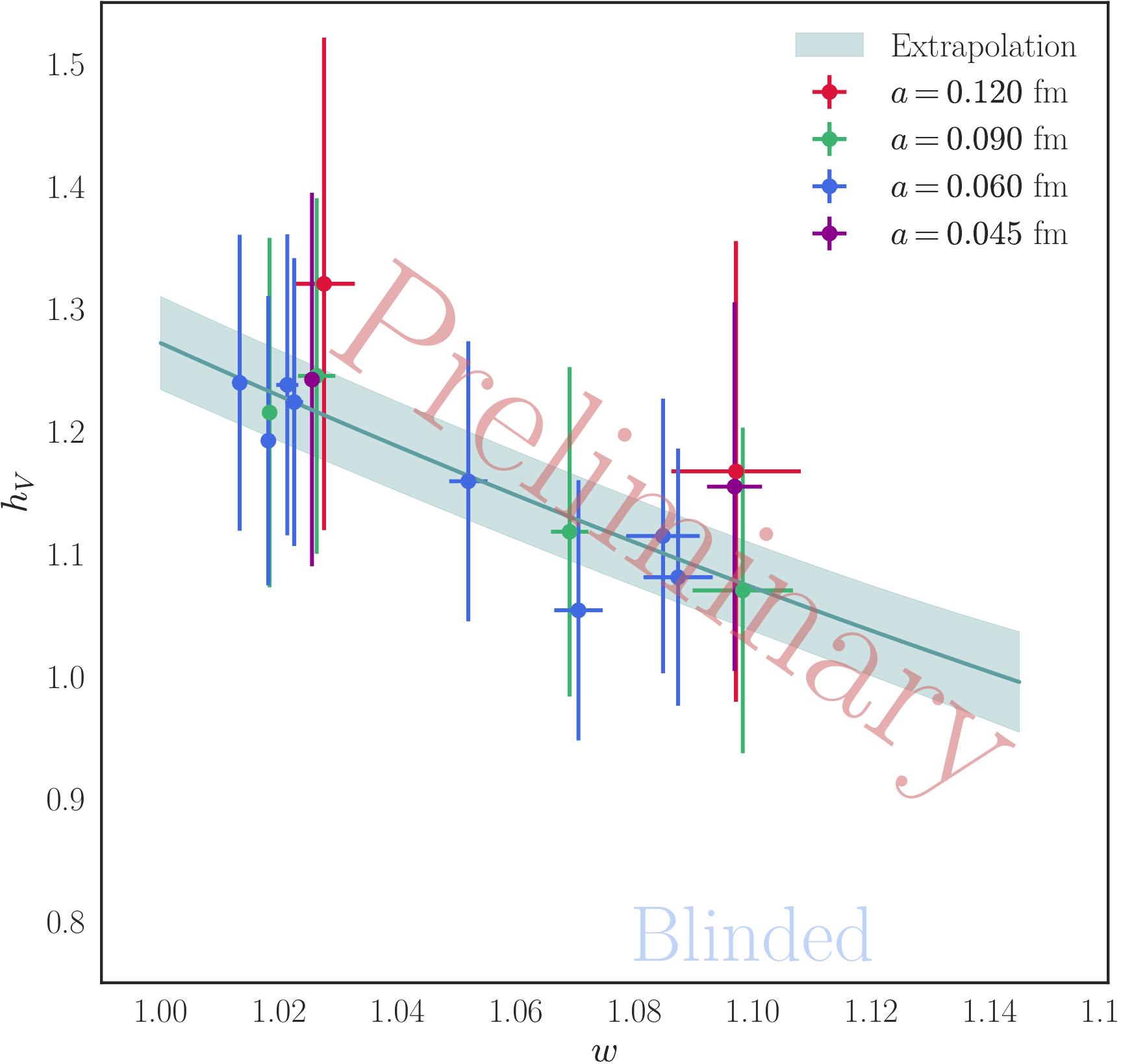} }
  \subfigure[$h_{A_1}(w)$ form factor.]
            {\includegraphics[width=0.40\linewidth,angle=0]{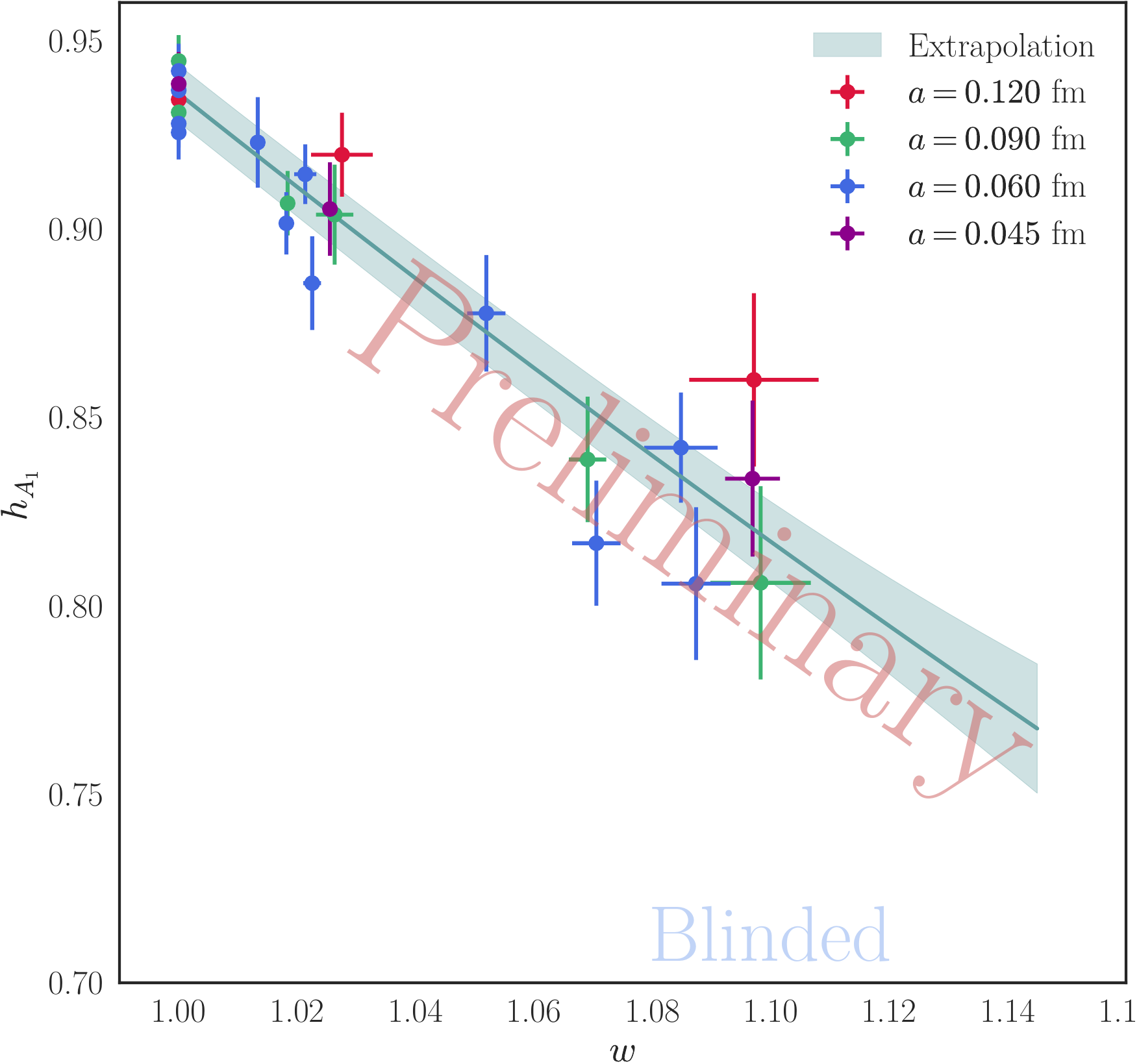} } \\
  \subfigure[$h_{A_2}(w)$ form factor.]
            {\includegraphics[width=0.40\linewidth,angle=0]{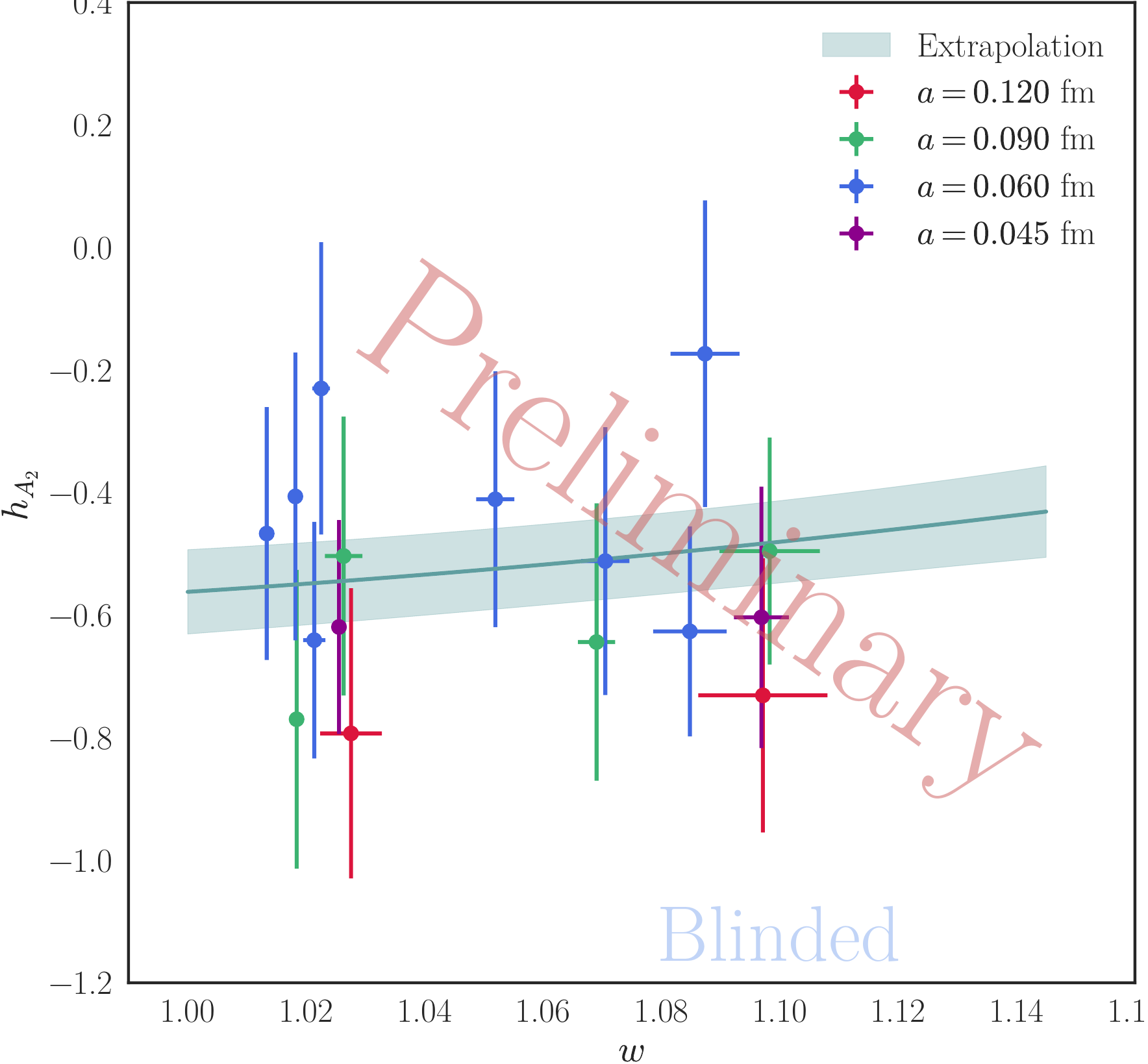} }
  \subfigure[$h_{A_3}(w)$ form factor.]
            {\includegraphics[width=0.40\linewidth,angle=0]{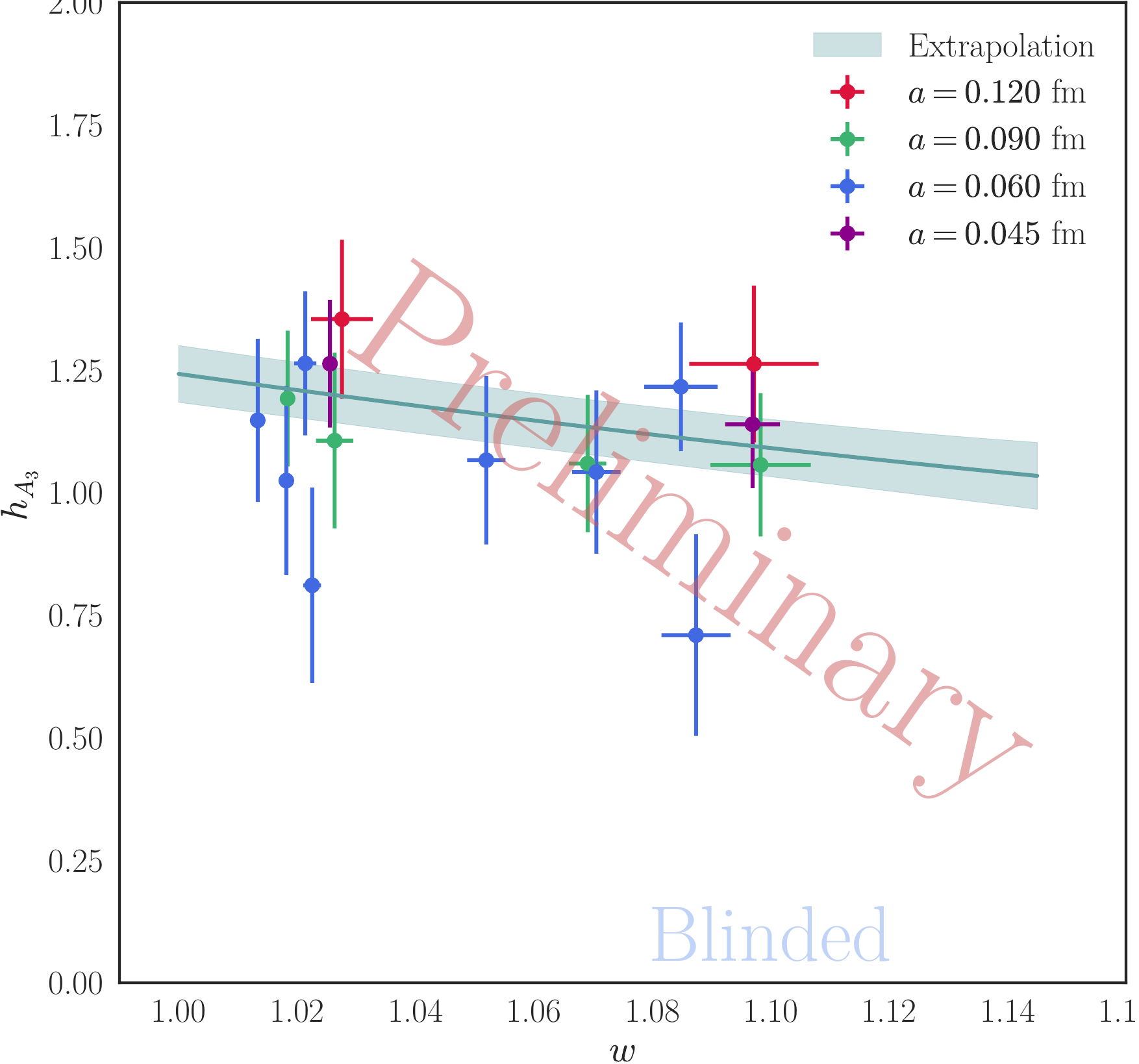} }
  \caption{Preliminary results for $h_V(w)$ and $h_{A_1}(w)$ in the upper row, and $h_{A_2}(w)$ and $h_{A_3}(w)$ in the lower row. The points are the lattice data for different lattice spacings, light quark
           masses and volumes, whereas the band represents the result of the chiral-continuum fit. The $p$-value of the joint fit is $p=0.54$.\label{fFact}}
\end{figure}
The ansatz for each form factor is given by
\begin{align}
h_X = 1& + \frac{X_{h_X}(\Lambda_{\textrm{QCD}})}{m_c} + \frac{g_{D^\ast D\pi}}{48\pi^2f^2_\pi r_1^2}\textrm{logs}_{\textrm{SU(3)}}(w,m_l,m_s,\Lambda_{\textrm{QCD}}) - \rho^2(w-1) + k(w-1)^2 \label{ChCtFit} \\
       & + c_1x_l + c_2x_l^2 + c_{a_1}x_{a^2} + c_{a_2}x_{a^2}^2 + c_{a,m}x_lx_{a^2} + \sum_i z_{X,i}(a\Lambda_{\textrm{QCD}})f_{X,i}(m_0a), \nonumber
\end{align}
where $x_l = B_0 m_l / \left(2\pi f_\pi\right)^2$, $x_{a^2} = a^2 / \left(4\pi f_\pi r_1^2\right)^2$, $X_{h_X}(\Lambda_{\textrm{QCD}})$ is a function of $\Lambda_{\textrm{QCD}}$ that depends on the form
factor, and the ast term $\sum_i z_{X,i} f_{X,i}$ takes into account the discretization errors coming from the heavy currents, as described in~\cite{PhysRevD.85.114506}. For $h_{A_1}$ Luke's theorem states
that the second term on the rhs of Eq.~\eqref{ChCtFit} becomes $X_{h_{A_1}}(\Lambda_{\textrm{QCD}})/m_c^2$, and $h_{A_2}$ is normalized to zero at tree level.

We perform a simultaneous, correlated fit to the complete set of form factors $h_X$. Figure~\ref{fFact} shows the preliminary results of the chiral-continuum extrapolation for the four form factors. The
discretization errors for both the light and the heavy sector are now taken into account in the chiral-continuum extrapolation, and as they seem to account for a large percentage of the final error, the
coarsest ensembles have been omitted in the fit. Table~\ref{myTable} gathers a very preliminary error budget showing the most important contributions to the error of each form factor.

\begin{table}[h]
  \footnotesize
  \begin{center}
    \begin{tabular}{l|c|c|c|c}
      Source                            &  $h_V\,(\%)$   & $h_{A_1}\,(\%)$ & $h_{A_2}\,(\%)$ & $h_{A_3}\,(\%)$ \\
      \hline
      Statistics                        &     $1.1$      &      $0.4$      &      $4.9$      &       $1.9$     \\
 {\bf $\chi$PT/cont. extrapolation}     & $\mathbf{1.9}$ & $\mathbf{0.7}$  & $\mathbf{6.3}$  & $\mathbf{2.9}$  \\
 \emph{Matching}                        &   \emph{1.5}   &    \emph{0.4}   &   \emph{0.1}    &   \emph{1.5}    \\
 \emph{Heavy quark discretization}      &   \emph{2.5}   &    \emph{1.2}   &   \emph{9.0}    &   \emph{6.0}
    \end{tabular}
  \end{center}
%  \scriptsize
%  $^*$Estimate, analysis in progress
  \caption{Preliminary error budget. In bold the contribution to the error budget that will be reduced in the next step of our program, by using HISQ fermions at the physical masses in the light sector. In
           italic the part of the error that will be reduced when we introduce HISQ in the heavy sector as a second step in our roadmap.\label{myTable}}
\end{table}

\section{$z$ expansion}
\begin{figure}[h]
  \centering
  \subfigure[Separate fits]
            {\includegraphics[width=0.40\linewidth,angle=0]{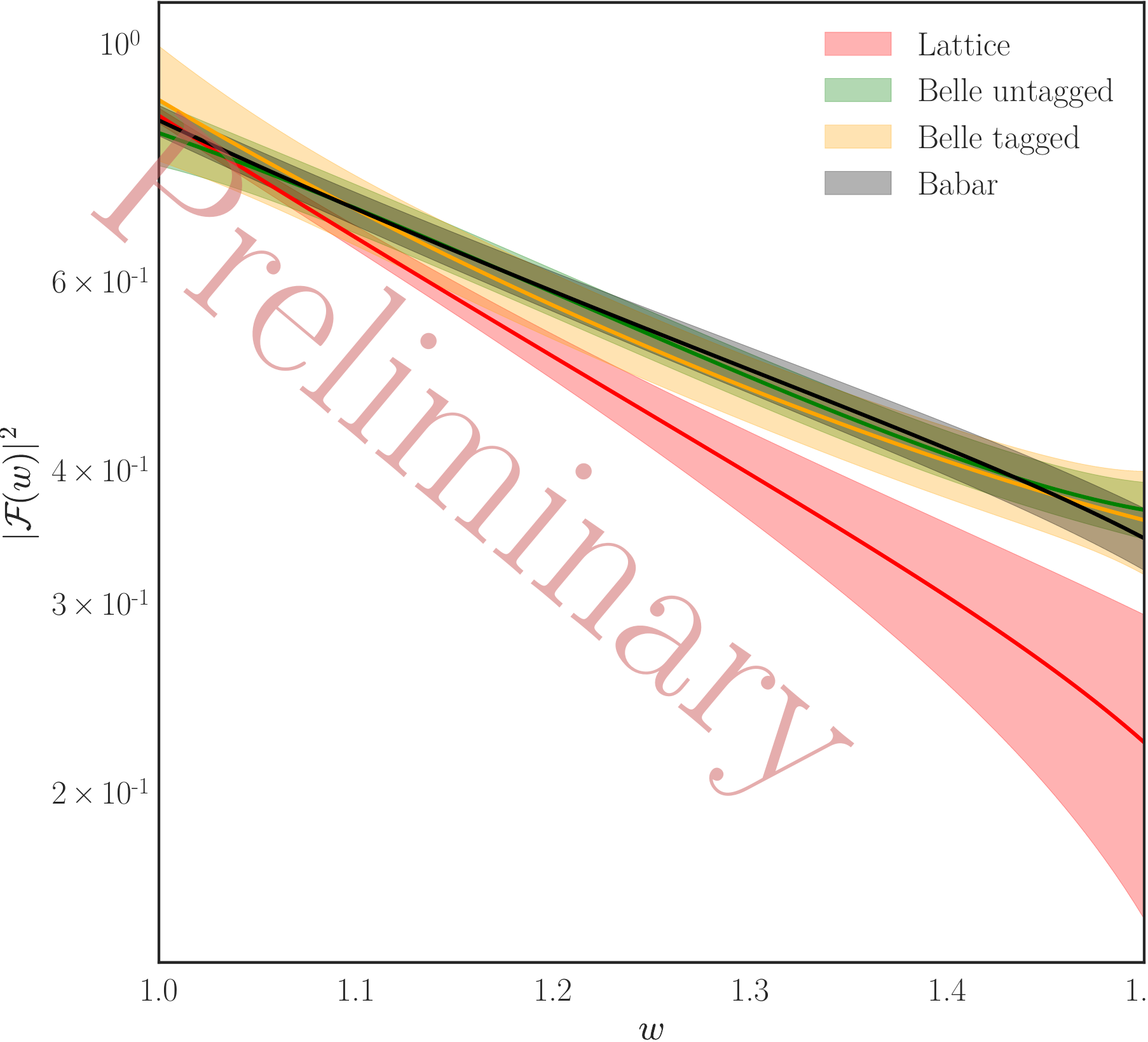} }
  \subfigure[Joint fit]
            {\includegraphics[width=0.40\linewidth,angle=0]{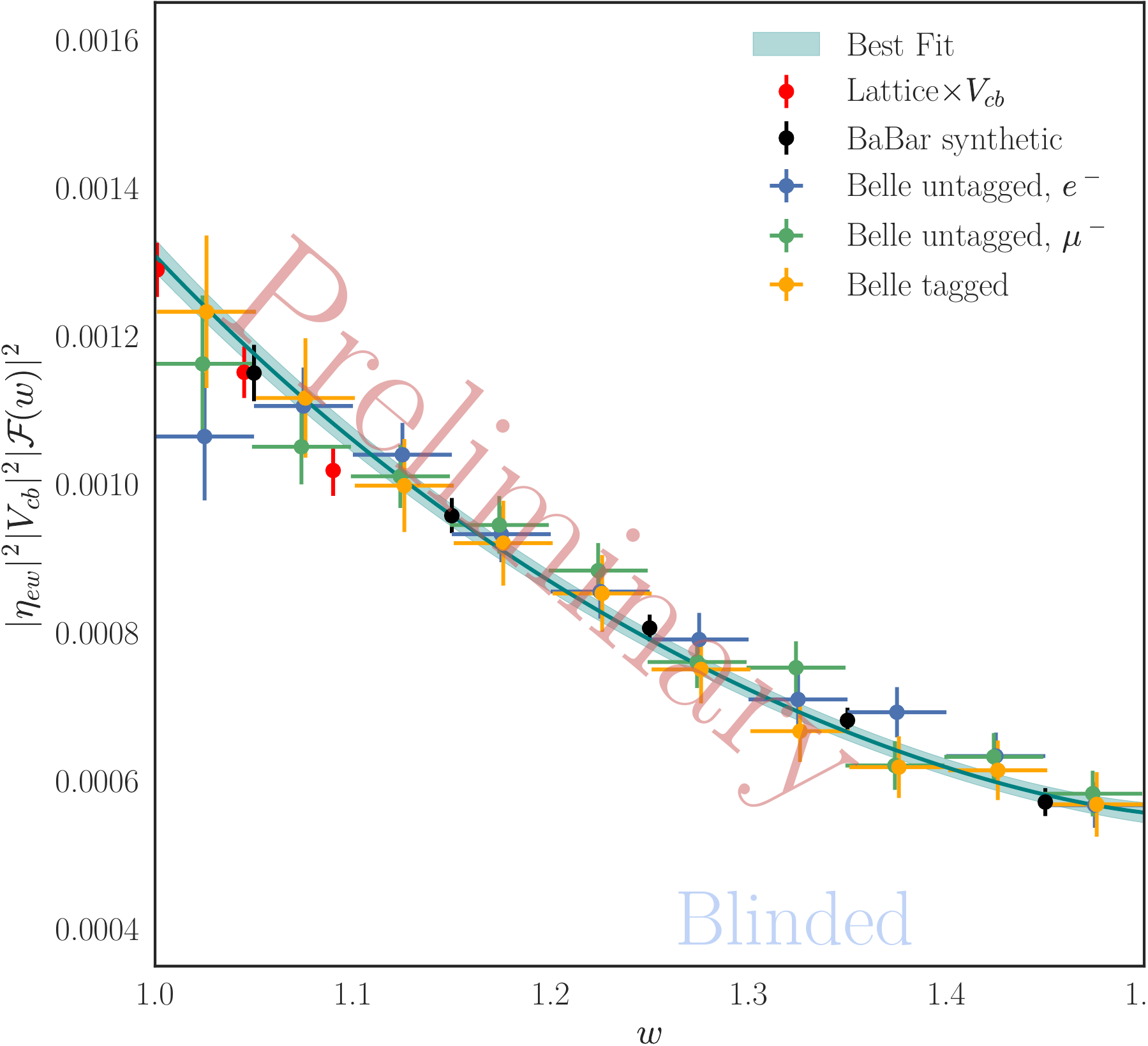}}
  \caption{Preliminary results for the $z$ expansion fits to lattice and experimental data. On the left, results for individual fits to the separate datasets. The $p$-value of the lattice-only fit is
           $p=0.35$. On the right, the joinmt fit of all the datasets together, with $p=0.29$.\label{zFits}}
\end{figure}

We use the BGL parametrization~\cite{Boyd:1997kz} to perform a fit using synthetic data constructed from the output of the chiral-continuum fit. We also add to the mix the available experimental results from
the Belle~\cite{BelleTagged,BelleUntagged} and BaBar \cite{BaBar} collaborations. In the left pane of Fig.~\ref{zFits} we perform separate fits of the four datasets (lattice data, Belle tagged, Belle untagged
and BaBar), and the right pane shows a joint fit of all the available data. Each dataset is fit using a different procedure: for the lattice we fit directly the BGL form factors to our data; in the Belle tagged
dataset the four different one-dimensional binnings ($w$ and the three angles) are fit at the same time; the Belle untagged dataset is fit using the same procedure as in~\cite{BelleUntagged}; and finally the
BaBar collaboration gives the fit results with correlations, so we simply generate synthetic data from their fit results.
The lattice analysis shows a larger slope at small recoil than the experimental measurements, which affects the decay amplitude at large recoil. We are investigating the source of this behavior.

\section{Preliminary conclusions}
Our unfinished analysis suggests caution before reaching any conclusions. In particular, we still need to finalize our analysis of the heavy-quark and recoil dependent discretization errors. Compared to current
determinations of $\left|V_{cb}\right|$ from experimental measurements of exclusive $B\to D^\ast$ decay rates together with the CLN parameterization and a lattice QCD input at only zero recoil, the results of
our analysis may or may not reduce the total uncertainty on $\left|V_{cb}\right|$. However, including information on the form factors away from zero recoil will make the $\left|V_{cb}\right|$ determination more
robust, apart from the extra information coming from the behavior of the form factors at small recoil. We have also designed a future roadmap to improve further the current calculation. In our present analysis
we have not used the CLN parametrization. Recent works~\cite{BelleUntagged,BaBar,JLQCD} suggest that the CLN parametrization is still reliable at the current error levels. In the final analysis we plan to
include a comparison between our CLN and our BGL results.

\section*{Acknowledgments}
Computations for this work were carried out with resources provided by the USQCD Collaboration, the National Energy Research Scientific Computing Center and the Argonne Leadership Computing Facility, which
are funded by the Office of Science of the U.S. Department of Energy; and with resources provided by the National Institute for Computational Science and the Texas Advanced Computing Center, which are funded
through the National Science Foundation's Teragrid/XSEDE Program. This work was supported in part by the U.S. Department of Energy under grant No. DE-SC0015655 (A.X.K.), by the U.S. National Science Foundation
under grants PHY10-67881 and PHY14-17805 (J.L.), PHY14-14614 and PHY17-19626 (C.D., A.V.); by the Fermilab Distinguished Scholars program (A.X.K.); by the German Excellence Initiative and the European Union
Seventh Framework Program under grant agreement No. 291763 as well as the European Union's Marie Curie COFUND program (A.S.K.). Fermilab is operated by Fermi Research Alliance, LLC, under Contract No.
DE-AC02-07CH11359 with the United States Department of Energy, Office of Science, Office of High Energy Physics.

\bibliographystyle{JHEP}
\bibliography{PoSLat19}

\end{document}